\documentclass[letterpaper, 10 pt, conference]{ieeeconf}    

\IEEEoverridecommandlockouts                              
\usepackage{graphics} 
\usepackage{epsfig} 
\usepackage{mathptmx} 
\usepackage{times} 
\usepackage{amsmath} 
\usepackage{amssymb}  
\usepackage[T1]{fontenc}
\usepackage{color}
\usepackage{soul}
\usepackage{caption}
\usepackage{subcaption}

\title{\LARGE \bf
Medically Relevant Criteria used in EEG Compression for Improved
Post-Compression Seizure Detection}

\author{Hoda Daou$^{1}$ and Fabrice Labeau$^{1}$
\thanks{This work was supported by the Natural Sciences and Engineering Research Council (NSERC) and industrial and government partners, through the Healthcare Support through Information Technology Enhancements (hSITE) Strategic Research Network.}
\thanks{$^{1}$ F. Labeau and H. Daou are with the Department of Electrical and Computer Engineering at McGill University, Montreal, QC H3A 2A7, CANADA {\tt\small fabrice.labeau at mcgill.ca, hoda.daou at mail.mcgill.ca.}}
}

\begin{document}

\maketitle
\thispagestyle{empty}
\pagestyle{empty}

\begin{abstract}
Biomedical signals aid in the diagnosis of different disorders and abnormalities. When targeting lossy compression of such signals, the medically relevant information that lies within the data should maintain its accuracy and thus its reliability. 
In fact, signal models that are inspired by the bio-physical properties of the signals at hand allow for a compression that preserves more naturally the clinically significant features of these signals. 
In this paper, we illustrate this through the example of EEG signals; more specifically, we analyze three specific lossy EEG compression schemes.  These schemes are based on signal models that have different degrees of reliance on signal production and physiological characteristics of EEG. The resilience of these schemes is illustrated through the performance of seizure detection post compression. 

\end{abstract}

\section{INTRODUCTION}


Nowadays medical information management systems and transmission of biomedical signals  are widely used in hospitals and clinics. In addition, transmission of biomedical signals allows medical experts to remotely evaluate the information carried by the signals in a cost-effective manner.
The massive amount of data requires large storage space and channel bandwidth and therefore, this problem calls for efficient compression methods.

There is a need to efficiently compress biomedical signals while preserving the important diagnostic-oriented information that lies within this data. Lossless compression guarantees no added distortion and therefore the data remains reliable for medical analysis. Although lossless compression is more desired for medical signals, higher compression rates can be achieved using lossy techniques.

When targeting lossy compression for biomedical signals, more focus should be given on retaining medically relevant information.  And thus, when coding the signals, more emphasis should be given to this particular aspect of the data in order to achieve good compression performance. {In this paper, taking the example of Electroencephaogaphy (EEG) signals, we argue that biomedical signal compression systems that take into account  more effectively the underlying nature of the signal lead to better results, in terms of the preservation of clinically significant features after compression.} In the next paragraphs we will focus on the characteristics of EEG signals, more specifically, on the underlying generators and the different neurological aspects of these signals. 

%


Most observed scalp EEG activity is generated within the cerebral cortex \cite{ica_zhukov}. A synchronous synaptic simulation of a very large number of neurons results in a dipolar current source oriented orthogonal to the cortical surface \cite{ica_zhukov}. The measured EEG is actually the propagation of this current onto the different electrodes' locations.

Thus, EEG signals can be considered as projections of certain activities that are occurring inside the cerebral cortex. These projected electrical signals are measured from certain locations on the patient's head, i.e. electrodes. Since certain neurological components are behind these observations, a lot of redundancy is present. This redundancy can be directly seen between the different recording channels. This is known as spatial redundancy. 

EEG is used to diagnose certain disorders and also in sleep analysis, known as polysomnography. These signals reflect the state of the patient. In fact, the different functional stages of a patient's state of mind can be characterized by certain EEG rhythms or brain waves \cite{journal}. Brain activity of EEG signals is usually  divided into five main frequency rhythms: delta ($0$ - $4$ Hz), theta ($4$ - $8$ Hz), alpha ($8$ - $13$ Hz), beta ($13$ - $30$ Hz) and gamma ($30$ - $100$ Hz) \cite[p. 33]{EEGB} \cite{journal}. The presence or absence of these waves during certain periods of recording can help determine certain abnormalities. 
Since these rhythms tend to naturally extend and repeat during different stages of the EEG recording,  there is redundancy present at certain frequency sub-bands between different periods of recording. 
 
When compressing these signals, the neurological characteristics that are usually used in the medical analysis of these signals can help achieve better analysis and approximation and thus better remove redundant information. 

We recently suggested three different methods that target the compression of scalp EEG data using different modelling and coding techniques.  These methods were developed while focusing on the neurological characteristics of these signals.
 
The first method is based on using classic transformation and coding techniques to compress the EEG recordings while focusing on spatial redundancy \cite{Hoda1}. The second method explores a common physiological characteristic of the EEG signals, more specifically brain waves, in order to develop appropriate compression methods. It focuses on extracting the redundancy present at specific frequency bands to achieve decorrelation at different time instances \cite{journal}. In the third method, the underlying physiological sources behind the observed signals on the scalp are explored. The observed signals are modelled using these sources which help in extracting the mutual information present between the EEG channels \cite{journal2}.

In this paper, these three different methods are first presented. Afterwards, performance results in terms of post compression seizure detection are shown. This recently proposed qualitative measure that is used to compare the original and reconstructed signals provides a better reflection of the information loss with respect to the medically relevant data  \cite{mypaper} \cite{journal}. A comparative analysis that discusses the weaknesses and strengths of each suggested method is then presented. The paper ends with a conclusion and suggestions for future work.

\section{Compression Methods}
\subsection{Pre-{P}rocessing of {M}ulti-{C}hannel {EEG} for {I}mproved {C}ompression {P}erformance using {SPIHT}}

As previously mentioned, in scalp recordings, EEG signals measured from certain locations on the scalp can be seen as the projection of activity located inside the brain \cite{Hoda1}. 
In fact, EEG channels display a lot of similarity and even superposition of the different signals. Thus looking at these recordings in the spatial dimension, i.e. between different channels, is very important in capturing this redundancy \cite{Hoda1}.

The first method uses discrete wavelet transform (DWT) and SPIHT in $2$D to code the EEG channels \cite{spiht1} \cite{  Lu99waveletcompression} \cite{Hoda1}. Thus, it makes use of the inter-channel redundancy present between different EEG channels of the same recording and the intra-channel redundancy between the different samples of a specific channel.  SPIHT was originally suggested for the compression of $2$D images, thus this method exploits the basic characteristics of this type of data. More precisely, it exploits images characteristics where most of the image's  energy is located in the low frequency components and there is  spatial self-similarity among the sub-bands \cite{Hoda1}.

In this SPIHT-based method, classic compression techniques that are initially targeted for $2$D images are applied on matrices of EEG recordings. However, pre-processing is performed as first step in order to optimize the performance of these coders on the characteristics of our signals \cite{Hoda1}. 

\subsection{Dynamic {D}ictionary for {C}ombined {EEG} {C}ompression and {S}eizure {D}etection}

When analyzing EEG signals for the purpose of medical diagnosis, brain waves are identified in order to find the different functional stages of a patient's state of mind. As previously mentioned, these different rhythms can be used to characterize the different EEG segments \cite{journal}. Thus, depending on the state of mind of the patient, brain waves tend to extend and repeat throughout different segments of recording. This creates redundancy between the segments.




The second suggested method, dictionary-based method, aims at comparing EEG segments of different time periods and extracting the redundancy present between these segments. To do that, this technique focuses on the energy in the different frequency sub-bands that correspond to the different brain rhythms. DWT, dynamic reference lists and SPIHT are used to compute and code the decorrelated sub-band coefficients. This method is able to both compress EEG channels and detect seizure-like activity \cite{journal}. 


Therefore, this method uses a physiological characteristic of EEG signals, which is the different brain waves, in order to analyze the signals and remove the intrinsic redundancy between the different segments in a single EEG channel.

\subsection{{EEG} {C}ompression of {S}calp {R}ecordings based on {D}ipole {F}itting}

As previously mentioned, there are certain neuronal generators that are behind the observed EEG signals \cite{review} \cite{journal2}. In fact, the non-invasive localization of these generators is known as the \emph{inverse solution} and is used in the medical analysis of EEG. Finding a solution to the inverse problem by relying on the pattern of recorded EEG is able to give us a model that maps the generators to the measured projections on the scalp \cite{review}. Therefore, having solved the inverse problem, one can use such a model to generate, from the calculated dipoles, an approximation of the EEG recordings. This is known as the forward problem \cite{journal2}.


This third method, dipole-based method, provides a deeper analysis of the intrinsic dependency inherent between the different EEG channels. It is based on dipole fitting that is usually used in order to find a solution to the classic problems in EEG analysis: inverse and forward problems \cite{review} \cite{forwp1}. The suggested compression system uses dipole fitting as a first building block to provide an approximation of the recorded signals. Then, based on a smoothness factor, appropriate coding techniques are suggested to compress the residuals of the fitting process. 

\section{Results and Discussions}
   As previously mentioned, in medical signals, it is important to move towards a diagnostics-oriented performance assessment \cite{mypaper} \cite{journal}. In the next section we will focus on analyzing the performance of the three suggested compression methods using an automatic seizure detection system Stellate Harmonie System \cite{mypaper} \cite{journal}.
   
   \subsection{Dataset}
   Data used in the testing, known as CHB-MIT Scalp EEG Database, was collected at the Children's Hospital Boston \cite{physiobank}. Recordings are done on pediatric patients suffering from intractable seizures. These recordings are annotated by medical experts and are sampled at $256$ Hz and $16$ bits used in the recording's precision. 


\begin{table*}

\small
\caption{Detection results with Stellate Harmonie tested at $16$ bps as Ground Truth.}

\begin{center}
\begin{tabular}{ccccccccccccc}
      & \multicolumn{4}{c}{Dipole-Based}  & \multicolumn{4}{c}{1D Dictionary-Based} & \multicolumn{4}{c}{2D SPIHT-Based}  \\ 	 \cline{2-13} 
     &\multicolumn{2}{c}{2 bps} &\multicolumn{2}{c}{4 bps} &\multicolumn{2}{c}{2 bps} &\multicolumn{2}{c}{4 bps} &\multicolumn{2}{c}{2 bps} &\multicolumn{2}{c}{4 bps} \\ 	 
&  $TP$ (\%)  & $FP$    & $TP$ (\%) 		& 			$FP$&   $TP$ (\%)  & $FP$    & $TP$ (\%) 		& 			$FP$ &   $TP$ (\%)  & $FP$    & $TP$ (\%) 		& 			$FP$\\
  Detections &  & Count   & 		& 	Count &   & Count  & 		& 	Count & & Count &  		& 			Count\\
     $	6	$&$	83.33	$&$	1	$&$	100	$&$	0	$&$	100	$&$	0	$&$	83.33	$&$	1	$&$	83.33	$&$	1	$&$	100	$&$	0	$\\
$	1	$&$	100	$&$	0	$&$	100	$&$	0	$&$	100	$&$	0	$&$	100	$&$	0	$&$	100	$&$	0	$&$	100	$&$	0	$\\
$	9	$&$	88.89	$&$	1	$&$	100	$&$	0	$&$	88.89	$&$	1	$&$	100	$&$	0	$&$	88.89	$&$	1	$&$	88.89	$&$	1	$\\
$	6	$&$	100	$&$	0	$&$	50	$&$	3	$&$	33.33	$&$	4	$&$	50	$&$	3	$&$	16.67	$&$	5	$&$	0	$&$	6	$\\
$	8	$&$	87.5	$&$	1	$&$	87.5	$&$	1	$&$	100	$&$	0	$&$	100	$&$	0	$&$	100	$&$	0	$&$	100	$&$	0	$\\
$	25	$&$	88	$&$	3	$&$	100	$&$	0	$&$	100	$&$	0	$&$	100	$&$	0	$&$	60	$&$	10	$&$	96	$&$	1	$\\
$	5	$&$	100	$&$	0	$&$	100	$&$	0	$&$	100	$&$	0	$&$	100	$&$	0	$&$	100	$&$	0	$&$	100	$&$	0	$\\
$	2	$&$	100	$&$	0	$&$	100	$&$	0	$&$	100	$&$	0	$&$	100	$&$	0	$&$	100	$&$	0	$&$	100	$&$	0	$\\
$	10	$&$	100	$&$	0	$&$	100	$&$	0	$&$	60	$&$	4	$&$	100	$&$	0	$&$	90	$&$	1	$&$	80	$&$	2	$\\
$	2	$&$	100	$&$	0	$&$	100	$&$	0	$&$	100	$&$	0	$&$	100	$&$	0	$&$	100	$&$	0	$&$	100	$&$	0	$\\
$	4	$&$	100	$&$	0	$&$	75	$&$	1	$&$	75	$&$	1	$&$	75	$&$	1	$&$	75	$&$	1	$&$	100	$&$	0	$\\
$	9	$&$	77.78	$&$	2	$&$	100	$&$	0	$&$	77.78	$&$	2	$&$	88.89	$&$	1	$&$	77.78	$&$	2	$&$	88.89	$&$	1	$\\
\hline
Average	   &$	93.79	$&$	0.67	$&$	92.71	$&$	0.42	$&$	86.25	$&$	1.00	$&$	91.44	$&$	0.5	$&$	82.64	$&$	1.75	$&$	87.81	$&$	0.92	$	\\
\hline
\hline
\vspace{-0.3cm}
\end{tabular}
\end{center}
\end{table*}

\subsection{Statistical Measures used in Detection Analysis}

In order to test the performance of the different methods on the chosen datasets, the pre-processor of Stellate Harmonie, ICTA-S onset detector, is used. Testing is done with data compressed at different bit rates and flagged sections are compared in order to analyze the information loss. The statistical measures described below are similar to the ones explained in previous studies \cite{journal} \cite{mypaper}. 


The percentage of true positives ($TP$) and the total number of false positives ($FP$) are used in the evaluation process. In these measures, the ground truth is chosen from the detection output when testing the original EEG records. It is equal to the total number of flagged sections found. The following provides a definition of the statistical measures for this scenario \cite{mypaper} \cite{journal}:
\begin{itemize}
\item{True Positive ($TP$): A period of one minute or more of overlap occurs between a flagged section in the compressed file and a flagged section in the original file.
}
\item{False Positive ($FP$): No overlap, or an overlap of less than a minute, is found between a flagged section in the compressed recording and flagged sections in the original recording.
} 
\end{itemize}

The total number of $TP$ is divided by the total number of flagged sections in the original file in order to compute the percentage of the true positives. 

\subsection{Results}



Table $I$ shows the detection results of all three methods when taking Stellate Harmonie tested on the original files (i.e. at $16$ bps) as ground truth. When looking at both the individual patients and on the average values over all patients (shown in the last row), we notice that there is degradation in performance between the different methods. It can clearly be seen that the dipole-based method gives higher percentage of $TP$ for most patients compared to both methods. In addition, the dictionary-based method outperforms the $2$D SPIHT-based method. The same can be observed for the total number of false positives. The number of $FP$ for the dipole-based method is lower than the one for the dictionary-based method, which also is lower than the $2$D SPIHT-based method. This can be observed for most of the patients used in the testing.

\begin{figure}
\centering
\includegraphics[scale=0.85]{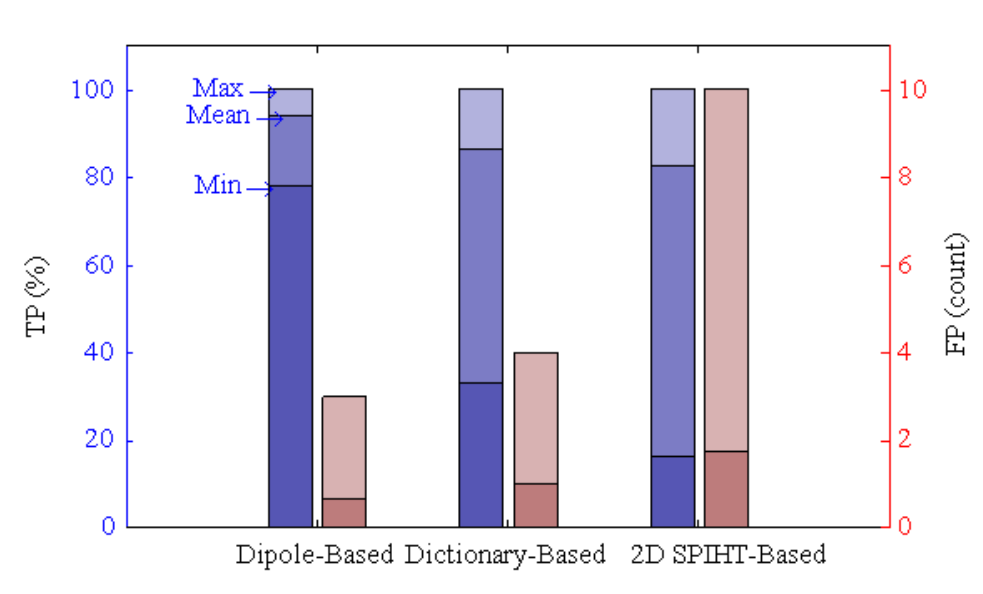}
\caption{Bar plots showing the mean, maximum and minimum values of the detection results with Stellate Harmonie tested at $16$ bps as Ground Truth and bit rate equal to $2$bps.}
 
\end{figure}

\begin{figure}
\centering
\includegraphics[scale=0.85]{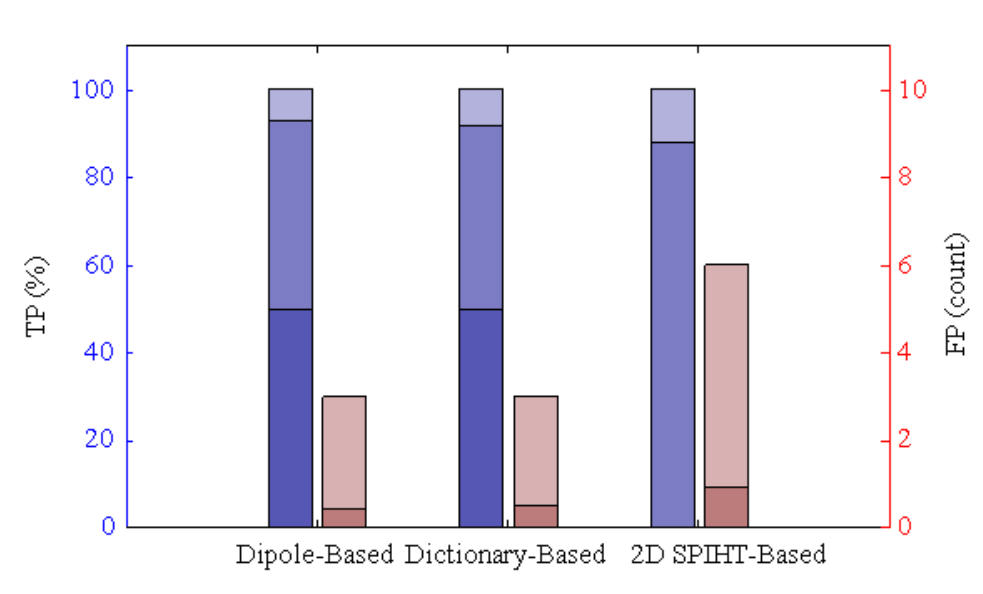}
\caption{Bar plots showing the mean, maximum and minimum values of the detection results with Stellate Harmonie tested at $16$ bps as Ground Truth and bit rate equal to $4$bps.}
 
\end{figure}

Figures $1$ and $2$ summarize the results shown in Table $I$ by highlighting the mean, minimum and maximum values of $TP$ and $FP$ for the two bit rates and for the three different methods. It should be noted that certain parameters have $0$ minimum values, for this reason certain bars only show two values, the average and the maximum.  

These figures highlight the fact that the mean and the minimum values of $TP$ are highest for the dipole-based method and decrease as we switch to the dictionary-based method then to the $2$D SPIHT-based method. In addition, all methods at all bit rates have a maximum value of $TP$ of $100\%$.  The opposite is noticed for the false positives where an increase occurs between these three methods. We actually notice a big jump in $FP$ for the $2$D SPIHT-based method for a bit rate of $4$ bps.

\begin{figure}
\centering
\begin{minipage}{0.5\textwidth}

\begin{subfigure}[a]{0.9\textwidth}

 \includegraphics[scale=0.75]{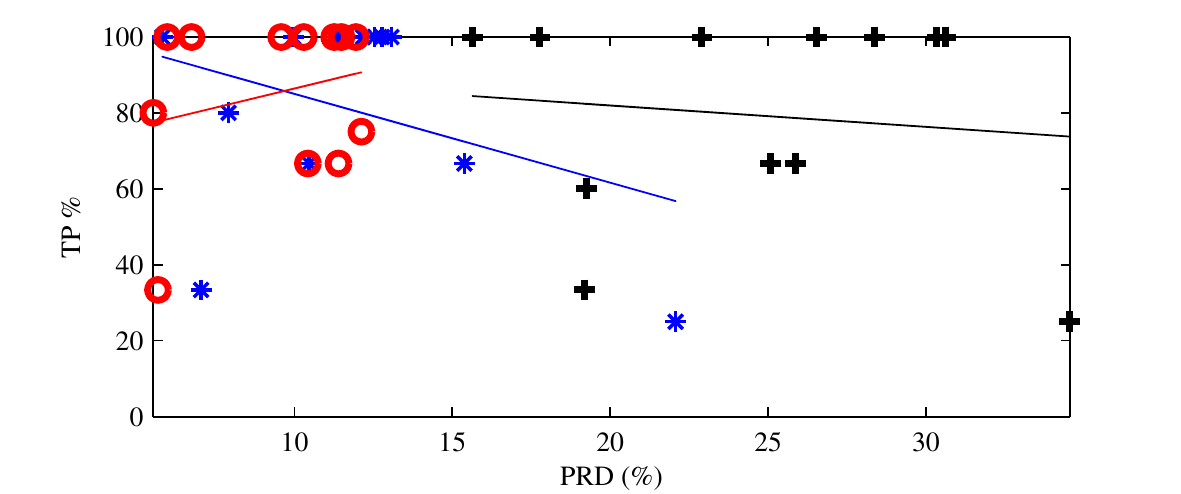} 
 \centering  
 \caption{Dipole-based ({{\color{red}o}}), dictionary-based (+) and $2$D SPIHT-based ({\color{black}{*}}) at $2$ bps.} 
 \end{subfigure}  
\end{minipage}
\begin{minipage}{0.5\textwidth}
\begin{subfigure}[b]{0.9\textwidth}

   \includegraphics[scale=0.75]{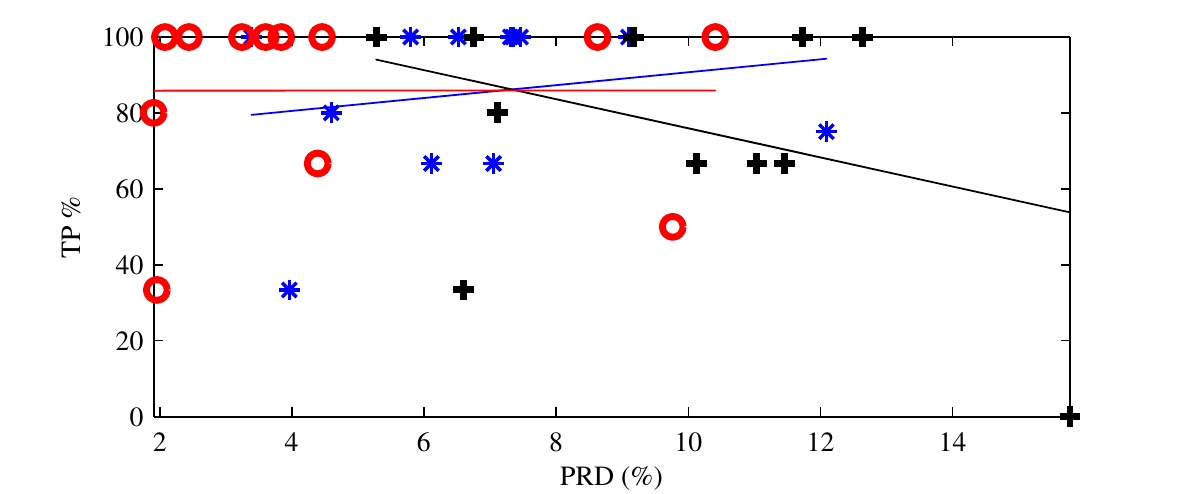}   
   \centering    
   \caption{Dipole-based ({{\color{red}o}}), dictionary-based (+) and $2$D SPIHT-based ({\color{black}{*}}) at $4$ bps.} 
   \end{subfigure}  
\end{minipage}
\vspace{10mm}
 \caption{Scatter plot showing $TP$ with the Stellate Harmonie detections of the original files as Ground Truth with respect to the mean PRD values of all $12$ patients of MIT DB.  }
\end{figure}


Figure $3$  shows the scatter plots of $TP$ where the ground truth is taken as the detections of Stellate Harmonie tested on the original files with respect to the mean PRD values of all $12$ patients of MIT DB, at bit rates $2$ and $4$ bps. Both Figures a and b show a direct relation between distortion and  true detections for the dictionary-based and the $2$D SPIHT-based methods. For these two methods, PRD values vary a lot and detections decrease as the values of PRD increase. However, for the dipole-based method, at both bit rates of $2$ and $4$ bps, PRD values do not vary a lot and all values of $TP$ are high. Thus there is no apparent relationship that links distortion and detections.


\subsection{Discussion}

In the $2$D SPIHT-based method, pre-processing transforms applied prior to coding focus more on the inter-channel redundancy which is a physiological characteristic of the EEG caused by the placement of electrodes in neighboring regions on the scalp. However, in this method, decorrelation does not go beyond a certain matrix. {\color{black}This means that this method does not examine redundancy present between different temporal sections of recording.} 

The second method takes into account a different physiological characteristic of the EEG, which is the presence of certain brain waves at certain periods of time. As mentioned previously, these brain rhythms are an indication of the patient's state of mind. 
Thus, in EEG recordings, different EEG segments can display similar features and characteristics. In fact, EEG segments can be grouped based on certain features for manual classification and abnormality detection \cite{agarwal_gotman_2001} \cite{journal}. In this method reference lists with dynamic update are used to achieve this grouping.


When comparing the first two methods, an improvement is observed in the detection results where we notice an increase in true detections and a decrease in false positives. 
In fact, the $2$D SPIHT-based method gives the worst results in terms of compression distortion and seizure detection. This method is based on $2$D SPIHT coding that uses a tree-like hierarchy. In this hierarchy, low frequency components are considered to have higher energy than high frequency components.  Thus, when using SPIHT coding, less and less bits are allocated to the high frequency components as bit rates decrease. This causes more distortion in the high frequency band. 

Seizures sometimes manifest an increase in amplitude and frequency. Distortion added to the high frequency sections causes a degradation in true detections. Thus, for this reason, the first suggested compression method is not recommended for recordings of patients suffering from epilepsy. This method is not able to well preserve important diagnostic oriented information when allocated bit rates decrease.

The second suggested method, dictionary-based method, gives good detection results compared to the first method. Average values of $TP$ are almost as good as the dipole-based method, as seen in Figures $1$ and $2$. This method is able to detect seizure-like activity as shown in \cite{journal}. Thus, compression based on this method is recommended for recordings of patients suffering with epilepsy, where detection and compression of the data can be performed in parallel. 

As mentioned previously, the third method, i.e. dipole-based method, is based on modelling the relationships between the different channels using dipoles and their moments. It examines and explores a deeper physiological characteristic of the EEG compared to the other two methods, which is the fact that the signals are generated by dipoles located inside the skull. 
Thus, it provides better extraction of the redundancy between the different channels. In addition the suggested coding techniques further decorrelate the EEG matrices in time. This improvement in coding is highlighted in the results shown in \cite{journal2}. The third method is able to provide both lower distortion values for high $CRs$ and improvement in seizure detection even at low bit rates compared to the other two compression methods.

\section{CONCLUSION}

Results show that when exploring physiological characteristics of EEG signals, better extraction of redundancy can be achieved. In addition, the deeper and more meaningful the physiological feature used, the better the compression. 

The $2$D SPIHT-based method uses basic decorrelation by relying on the spatial and temporal redundancies that characterize the EEG. It applies simple pre-processing techniques and $2$D transform and coder. Improvement is achieved in the dictionary-based method where dynamic reference lists enable us to examine and explore a more pronounced physiological characteristic of the EEG, which is the presence of brain waves. 
A deeper extraction of the redundancy present between the channels is achieved in the dipole-based method, where dipole fitting is used to model the relationship between these channels and therefore explore a deeper physiological characteristic of the EEG. The coders used in this method achieve further decorrelation in $2$D. 

Results highlight the improvements in performance achieved from the first suggested method, $2$D SPIHT-based method to the latest suggested method, the dipole-based method. When the method is able to achieve better decorrelation of the recorded signals, an improvement in post-compression detection performance is achieved for very low bit rates.

The dipole-based method is based on the assumption that a single dipole is behind the generation of the observed activity on the scalp. This gives very low distortion for event-related potentials \cite{journal2}. Improvements can be added to this method by exploring the usage of a larger number of dipoles for different types of EEG recordings.

\section*{ACKNOWLEDGMENT}
The authors gratefully acknowledge Professor Jean Gotman and his team at the Montreal Neurological Institute and Hospital, McGill University, for helping and allowing us to use Stellate Hamronie System.


%
%
%
%
%

\bibliographystyle{IEEEtran}
\bibliography{EEGBooks}

\begin{thebibliography}{10}
\providecommand{\url}[1]{#1}
\csname url@samestyle\endcsname
\providecommand{\newblock}{\relax}
\providecommand{\bibinfo}[2]{#2}
\providecommand{\BIBentrySTDinterwordspacing}{\spaceskip=0pt\relax}
\providecommand{\BIBentryALTinterwordstretchfactor}{4}
\providecommand{\BIBentryALTinterwordspacing}{\spaceskip=\fontdimen2\font plus
\BIBentryALTinterwordstretchfactor\fontdimen3\font minus
  \fontdimen4\font\relax}
\providecommand{\BIBforeignlanguage}[2]{{%
\expandafter\ifx\csname l@#1\endcsname\relax
\typeout{** WARNING: IEEEtran.bst: No hyphenation pattern has been}%
\typeout{** loaded for the language `#1'. Using the pattern for}%
\typeout{** the default language instead.}%
\else
\language=\csname l@#1\endcsname
\fi
#2}}
\providecommand{\BIBdecl}{\relax}
\BIBdecl

\bibitem{ica_zhukov}
L.~Zhukov, D.~Weinstein, and C.~Johnson, ``Independent {C}omponent {A}nalysis
  for {EEG} source localization,'' \emph{IEEE Engineering in Medicine and
  Biology Magazine}, vol.~19, no.~3, pp. 87 --96, 2000.

\bibitem{journal}
H.~Daou and F.~Labeau, ``Dynamic {D}ictionary for {C}ombined {EEG}
  {C}ompression and {S}eizure {D}etection,'' \emph{IEEE Journal of Biomedical
  and Health Informatics}, vol.~18, no.~1, pp. 247--256, January 2014.

\bibitem{EEGB}
E.~Niedermeyer and F.~Da~Silva, Eds., \emph{{E}lectroencephalography},
  5th~ed.\hskip 1em plus 0.5em minus 0.4em\relax Lippincott Williams and
  Wilkins, 2005, vol.~7.

\bibitem{Hoda1}
H.~Daou and F.~Labeau, ``Pre-{P}rocessing of {M}ulti-{C}hannel {EEG} for
  {I}mproved {C}ompression {P}erformance using {SPIHT},'' in \emph{Proceedings
  of the 34th Annual International Conference of the IEEE Engineering in
  Medicine and Biology Society (EMBS), San Diego, California USA, 28 August - 1
  September, 2012.}, 2012, pp. 2232 -- 2235.

\bibitem{journal2}
H.~Daou and F.~Labeau, ``{EEG} {C}ompression of {S}calp {R}ecordings based on {D}ipole
  {F}itting,'' \emph{IEEE Journal of Biomedical and Health Informatics},
  Submitted in November 2013, available at:
  \url{http://arxiv.org/abs/1403.2001}.

\bibitem{mypaper}
H.~Daou and F.~Labeau, ``Performance analysis of a $2$-{D} {EEG} {C}ompression {A}lgorithm
  using an {A}utomatic {S}eizure {D}etection {S}ystem,'' in \emph{Proceedings
  of Asilomar Conference on Signals, Systems, and Computers, Pacific Grove,
  California USA, 4 - 7 November 2012.}

\bibitem{spiht1}
D.~Rawat, C.~Singh, and M.~Sukadev, ``A hybrid coding scheme combining {SPIHT}
  and {SOFM} based vector quantization for effectual image compression,''
  \emph{European Journal of Scientific Research}, 2009.

\bibitem{Lu99waveletcompression}
Z.~Lu, Y.~Kim, Z.~Lu, D.~Y. Kim, and W.~Pearlman, ``Wavelet compression of
  {ECG} signals by the set partitioning in hierarchical trees ({SPIHT})
  algorithm,'' \emph{IEEE Transactions on Biomedical Engineering}, vol.~47, pp.
  849 -- 856, 1999.

\bibitem{review}
R.~Pascual-Marqui, ``{R}eview of {M}ethods for {S}olving the {EEG} {I}nverse
  {P}roblem,'' \emph{International Journal of Bioelectromagnetism}, vol.~1,
  no.~1, pp. 75 -- 86, 1999.

\bibitem{forwp1}
J.~Mosher, R.~Leahy, and P.~Lewis, ``{EEG} and {MEG}: forward solutions for
  inverse methods,'' \emph{IEEE Transactions on Biomedical Engineering},
  vol.~46, no.~3, pp. 245 --259, march 1999.

\bibitem{physiobank}
A.~L. Goldberger, L.~A.~N. Amaral, L.~Glass, J.~M. Hausdorff, P.~C. Ivanov,
  R.~G. Mark, J.~E. Mietus, G.~B. Moody, C.-K. Peng, , and H.~E. Stanley,
  ``Physiobank, physiotoolkit, and physionet: Components of a new research
  resource for complex physiologic signals,'' \emph{Circulation}, vol. 101,
  no.~23, p. 215 – 220, 2000.

\bibitem{agarwal_gotman_2001}
R.~Agarwal and J.~Gotman, ``Long-term {EEG} compression for intensive-care
  settings,'' \emph{Engineering In Medicine and Biology (IEEE)}, vol.~20,
  no.~5, pp. 23--29, 2001.

\end{thebibliography}

\end{document}